\documentclass[%
 reprint,
superscriptaddress,
 amsmath,amssymb,
 aps,
prl,
]{revtex4-1}
\usepackage{floatflt}
\usepackage{subfigure}
\usepackage{pdfpages}
\usepackage[section]{placeins}
\usepackage{mathtools}
\usepackage{graphicx}
\usepackage{dcolumn}
\usepackage{bm}
\makeatletter
\newcommand*{\balancecolsandclearpage}{%
  \close@column@grid
  \clearpage
  \twocolumngrid
  }
\makeatother

\begin{document}

\preprint{APS/123-QED}

\title{Surface and Step Conductivities on Si(111) Surfaces}

\author{Sven Just}
\affiliation{Peter Gr\"{u}nberg Institut (PGI-3) and
JARA-Fundamentals of Future Information Technology,
Forschungszentrum J\"{u}lich, 52425 J\"{u}lich, Germany}

\author{Marcus Blab}
\affiliation{Peter Gr\"{u}nberg Institut (PGI-3) and
JARA-Fundamentals of Future Information Technology,
Forschungszentrum J\"{u}lich, 52425 J\"{u}lich, Germany}

\author{Stefan Korte}
\affiliation{Peter Gr\"{u}nberg Institut (PGI-3) and
JARA-Fundamentals of Future Information Technology,
Forschungszentrum J\"{u}lich, 52425 J\"{u}lich, Germany}

\author{Vasily Cherepanov}
\affiliation{Peter Gr\"{u}nberg Institut (PGI-3) and
JARA-Fundamentals of Future Information Technology,
Forschungszentrum J\"{u}lich, 52425 J\"{u}lich, Germany}

\author{Helmut Soltner}
\affiliation{Central Institute of Engineering, Electronics and Analytics (ZEA-1), 
Forschungszentrum J\"{u}lich, 52425 J\"{u}lich, Germany}

\author{Bert Voigtl\"{a}nder}
\email[Corresponding author: ]{b.voigtlaender@fz-juelich.de}
\affiliation{Peter Gr\"{u}nberg Institut (PGI-3) and
JARA-Fundamentals of Future Information Technology,
Forschungszentrum J\"{u}lich, 52425 J\"{u}lich, Germany}




\date{\today}

\begin{abstract}
	Four-point measurements using a multi-tip scanning tunneling microscope (STM) are carried out in order to determine surface and step conductivities on Si(111) surfaces. In a first step, distance-dependent four-point measurements in the linear configuration are used in combination with an analytical three-layer model for charge transport to disentangle the 2D surface conductivity from non-surface contributions. A termination of the Si(111) surface with either Bi or H results in the two limiting cases of a pure 2D or 3D conductance, respectively. In order to further disentangle the surface conductivity of the step-free surface from the contribution due to atomic steps, a square four-probe configuration is applied as function of the rotation angle. In total this combined approach leads to an atomic step conductivity of $\sigma_\mathrm{step} = (29 \pm 9)$\,$\mathrm{\Omega}^{-1} \mathrm{m}^{-1}$ and to a step-free surface conductivity of $\sigma_\mathrm{surf} = (9 \pm 2) \cdot 10^{-6}\,\mathrm{\Omega}^{-1}/\square$ for the Si(111)-(7$\times$7) surface. 
\end{abstract}

\pacs{Valid PACS appear here}
\maketitle


The increasing importance of surface conductance compared to conductance through the bulk in modern nanoelectronic devices calls for a reliable determination of the surface conductivity in order to minimize the influence of undesired leakage currents on the device performance or to use surfaces as functional units. A model system for corresponding investigations is the Si(111)-(7$\times$7) 
surface. Over the years a wide range of values for the conductivity of this surface has been reported, spanning several orders of magnitude \cite{Wells1}, and the latest measurements deviate still by a factor of 2 to 3 \cite{Hasegawa1,Wolkow}. 
The main difficulty in measuring the surface conductivity is to separate the 2D conductance at the surface from the conductance through other channels, e.g. the bulk and the space charge layer.

\begin{figure}[bf]
\centering
\includegraphics[width=0.455\textwidth]{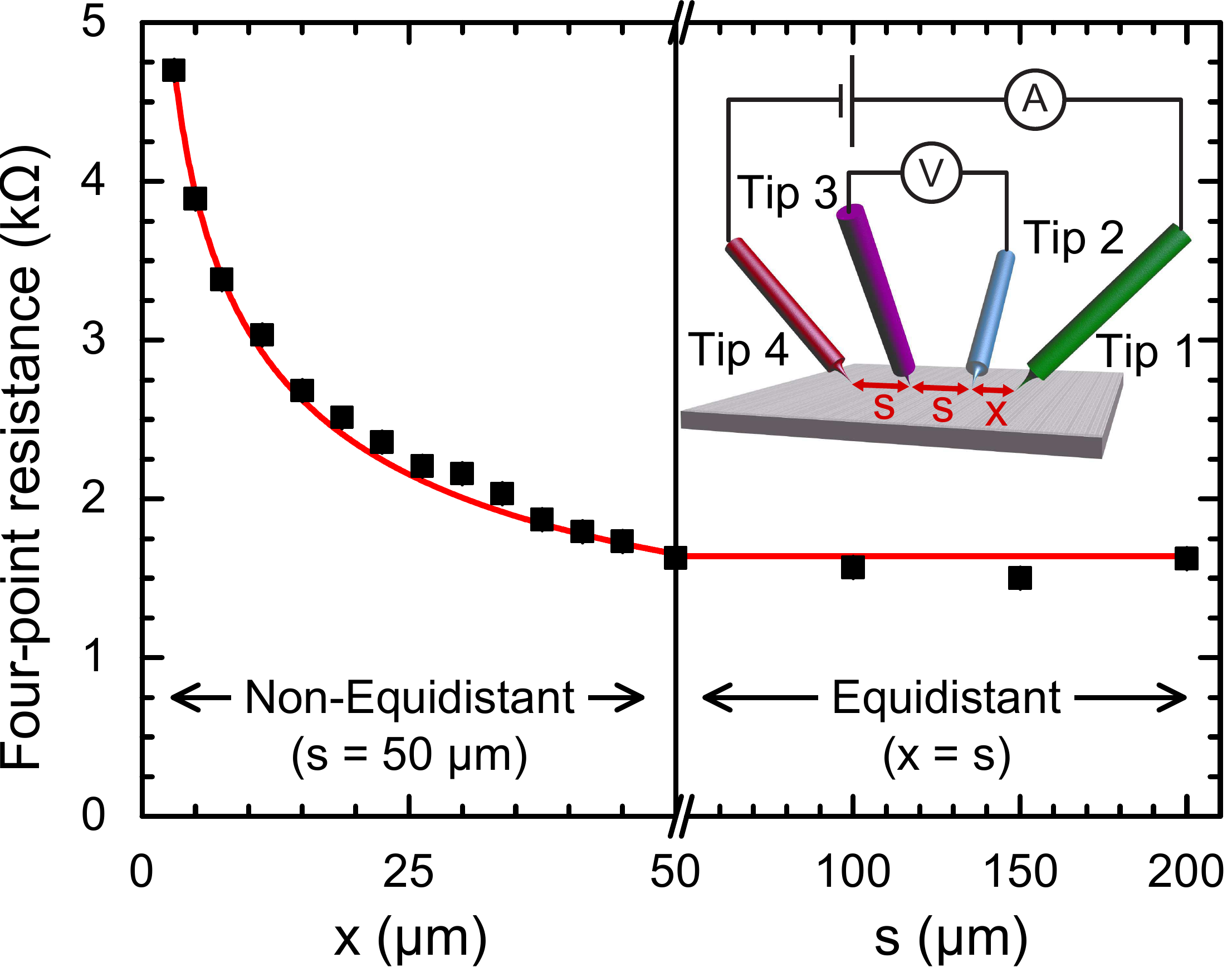}
\caption{(Color online) Four-point resistance of a Bi-terminated Si(111)-$(\sqrt{3} \times \sqrt{3})$R30$^\circ$ sample as function of the probe distances $s$ and $x$ for the equidistant (right half) and the non-equidistant configuration (left half). The red solid line represents the behavior expected for a pure 2D conductance with $\sigma_\mathrm{Bi} = (1.4 \pm 0.1)\cdot 10^{-4}\,\mathrm{\Omega}^{-1}/\square$. In the inset the linear measurement configuration is shown.}
\label{fig1}	
\end{figure}

Here, we use a four-tip scanning tunneling microscope \cite{4tipSTM} for distance-dependent measurements of the four-point resistance on Si(111), as shown in the inset in Fig.~\ref{fig1} for a linear tip arrangement, in combination with a three-layer model for charge transport. 
This method allows the separation of the surface conductance from other contributions due to the characteristic probe spacing dependency of different conductance channels.
Further on, we analyze the anisotropy of the surface conductance caused by the influence of atomic steps, which allows to determine the conductivity of a single step and the step-free surface. 

Analytic equations relating the measured four-point resistance to a conductivity can be obtained easily for pure 2D or 3D geometries, i.e. four tips positioned on a conducting sheet (surface) or on a half space (bulk), as \cite{Wells1} 
\begin{equation}\label{eq:1}
	R^{4p}_{2D}=\frac{\ln 2}{\pi \sigma_{2D}},\quad \mathrm{and}\quad R^{4p}_{3D}=\frac{1}{2 \pi \sigma_{3D}} \cdot s^{-1} 
\end{equation} 
with an equidistant probe spacing $s$, the 2D surface conductivity $\sigma_{2D}$ and the 3D bulk conductivity $\sigma_{3D}$. The equation for the 2D case shows the hallmark 
of a 2D channel, namely the fact that the surface conductance is independent of the probe spacing, while the conductance through a 3D channel depends on the distance $s$ of the four probes. 
In order to minimize the number of tips to be repositioned, we preferentially use a non-equidistant spacing, in which three tips remain at a mutual distance of $s = 50\,\mathrm{\mu m}$, while only the distance~$x$ between tip 1 and tip 2 is varied (Fig.~\ref{fig1}). In this non-equidistant setup the hallmark of the constant four-probe resistance is lost for the 2D case, since Eq.~\ref{eq:1} has to be modified as shown in \cite{Wells2,Wells3,Wojtaszek} and summarized in the Supplemental Material \cite{supmat}.

The four-point resistance measured on a Si(111)-$(\sqrt{3} \times \sqrt{3})$R30$^\circ$ Bi-terminated (1 ML) surface of an n-doped sample ($2\,\mathrm{k \Omega cm}$) is shown in Fig.~\ref{fig1} for the non-equidistant configuration with distances $x \le s = 50\,\mathrm{\mu m}$, and for the equidistant configuration with distances $x = s \ge 50\,\mathrm{\mu m}$
(details of sample preparation and measurement procedure are described in the Supplemental Material \cite{supmat}). 
The constant behavior in the equidistant range $s \ge 50\,\mathrm{\mu m}$ indicates a pure 2D character of conductance. Annother indicator for 2D surface transport is the fact that the four-point resistance, which is expected considering only the bulk conductivity, is several orders of magnitude larger than the observed one. 
Therefore, we compare the experimental data to a 2D model, and a good correspondence is obtained for $\sigma_\mathrm{Bi} = (1.4 \pm 0.1)\cdot 10^{-4}\,\mathrm{\Omega}^{-1}/\square$ (solid red line) confirming that the charge transport in the Bi-terminated Si(111) sample occurs almost exclusively through the 2D surface channel. Similar results were found for two differently doped samples. 

\begin{figure}[tbf]
\centering
\includegraphics[width=0.468\textwidth]{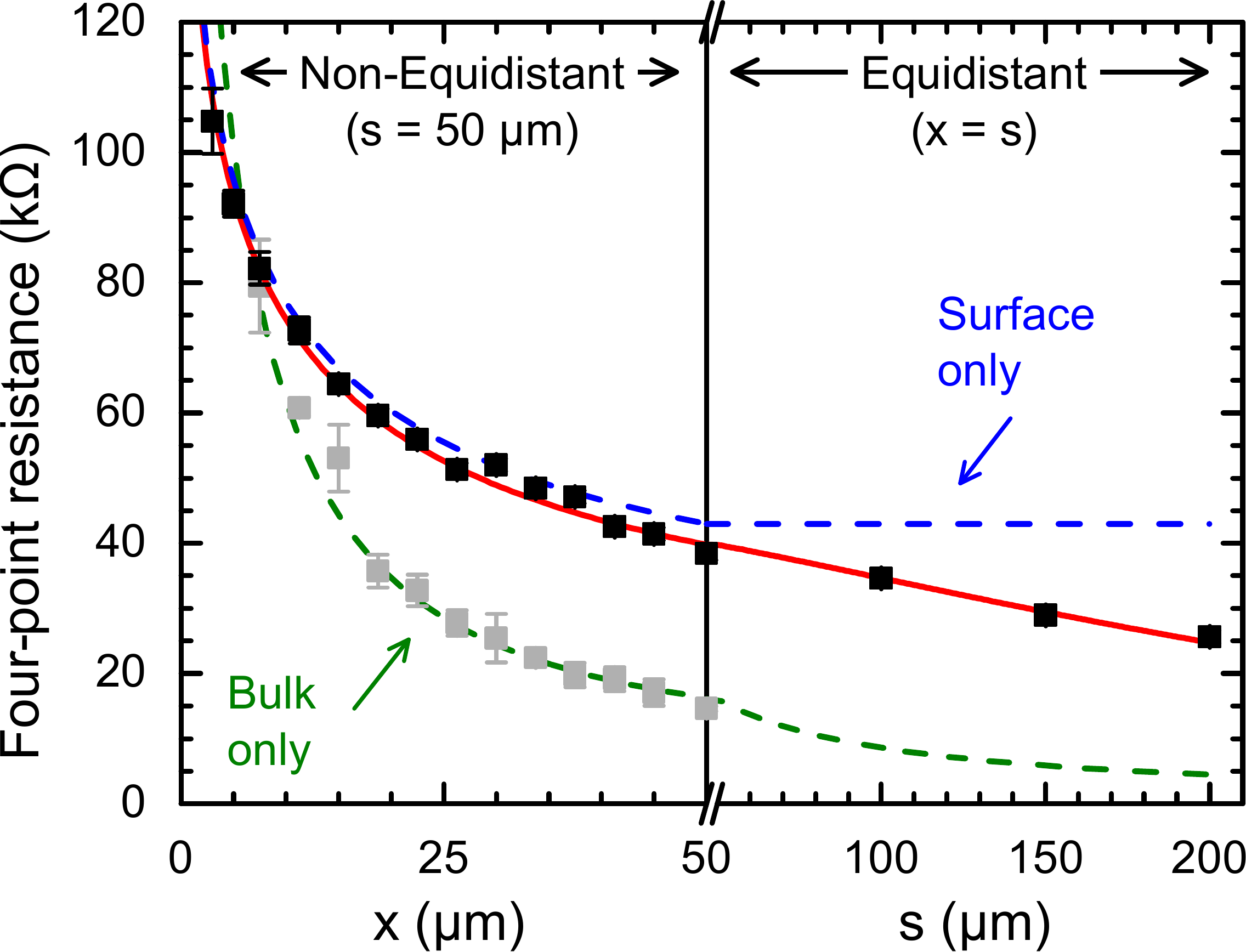}
\caption{(Color online) Four-point resistance of an n-doped Si(111)-(7$\times$7) sample as function of the probe distances $s$ and $x$ for the equidistant and the non-equidistant configuration. 
A three-layer model for charge transport yields the solid red line with $\sigma_{2D} = (5.1 \pm 0.7)\cdot 10^{-6}\,\mathrm{\Omega}^{-1}/\square$ located between the two limiting cases of a pure surface conductance (dashed blue line) and a pure bulk conductance (dashed green line). If the Si(111) surface is hydrogen-terminated, the surface contribution vanishes completely and the remaining bulk conductance can directly be measured (gray data points).}
\label{fig2}	
\end{figure}

Subsequently, the distance dependence of the four-point resistance was measured on a clean 
Si(111)-(7$\times$7) sample. The results for an n-doped sample ($700\,\mathrm{\Omega cm}$) are shown in Fig.~\ref{fig2}. 
The observed decreasing four-point resistance for increasing equidistant probe spacing $s$ indicates that a non-surface channel contributes to the charge transport, since a pure 2D conduction exhibits a constant behavior in the equidistant region (cf.~Fig.~\ref{fig1}). 
Thus, the measured four-point resistance should be modeled by a conductance channel through the surface states as well as additional contributions from the bulk and a near-surface space charge layer. However, in this case Eq.~\ref{eq:1} cannot be applied. 

Often an approximation of a parallel circuit consisting of the four-point resistance of the surface and the bulk (plus space charge layer) is used \cite{Wells3}, but this approach has two shortcomings. First, a complete separation of the surface conduction channel and the bulk is assumed. Second, the two-point resistances, not the four-point resistances, determine, which amount of current flows through the surface layer and which part through the bulk/space charge layer. So, the preferred way for the current to split up depends on the details of the injection, e.g. the size of the current injecting contact (tip diameter) \cite {Polley}. Thus, if more than one current path exists, the four-point resistance depends on possible transitions between charge transport channels as well as on the properties of the current injecting contacts, so that the well-known statement that the four-point resistance measured on the surface is independent of the contact resistances is not completely true.

In order to describe the charge transport through the different channels more accurately, we use an analytically derived three-layer model for conductance. 
In this model, the bulk enters with its known conductivity, while the surface conductivity is the parameter to be determined by a fit to the data. The space charge layer is approximated by an intermediate layer with a certain thickness and conductivity. These two values are obtained by the solution of the Poisson equation, which considers the known Fermi level pinning of the Si(111)-(7$\times$7) surface \cite{Lueth,Himpsel}. Nevertheless, the use of the bulk doping concentration as an initial parameter in this calculation in order to determine the width and average conductivity of the space charge layer is not sufficient to describe the data in Fig. \ref{fig2}.
However, it is known that high-temperature annealing up to 1200 $^{\circ}$C performed for cleaning the Si(111) surface causes a dopant redistribution and an additional p-type doping in the near-surface region due to boron in-diffusion \cite{Liehr,Iyer,Zhang,Robbins} or possible formation of near-surface single vacancies \cite{Bensalah}. These effects can lead to a reduced carrier concentration in the space charge layer. Generally, the details of the modification of the near-surface doping depend highly on the specific method and setup used for sample preparation. We find that the experimental data can be described well for a conductivity of $2.5 \cdot 10^{-4}\,\mathrm{\Omega}^{-1} \mathrm{m}^{-1}$ and a thickness of $3.1\,\mathrm{\mu m}$ for the intermediate layer representing the space charge layer. This quite approximate modeling of the space charge region as only one layer with constant conductivity seems to be sufficient, as the surface conductivity obtained from the fit to the measured data turns out to be insensitive to the specific properties of the intermediate layer. 

Overall, the three-layer model results in a much more accurate description of the measured four-point resistance than the simple parallel-circuit model, since it avoids the artificial separation between the surface and the non-surface channels and takes into account the injection geometry giving rise to a charge transport inside and between the layers according to their properties. 
The analytical derivation of the model is described in detail in the Supplemental Material \cite{supmat}.

The best fit to the measured four-point resistance using the three-layer model is shown as a solid red line in 
Fig.~\ref{fig2} and results in a surface conductivity of $\sigma_{2D} = (5.1 \pm 0.7)\cdot 10^{-6}\,\mathrm{\Omega}^{-1}/\square$. For comparison the two limiting cases are marked in Fig. \ref{fig2}: The four-point resistance arising from a pure 2D conductivity $\sigma_{2D}$ is shown as dashed blue line, while the four-point resistance induced by a pure 3D conductance, with its $1/s$ behavior in the equidistant configuration, is indicated as dashed green line featuring a bulk conductivity value, which is confirmed by an additional experiment described below. In the non-equidistant region the measured four-point resistance is close to the one expected from a pure surface conductance (less than 6\% deviation for $x \le 50\,\mathrm{\mu m}$), but for larger probe spacing an increasing deviation from the 2D behavior is observed. This reflects the well-known general tendency that the conductance is more surface-dominated for small probe distances, while a non-surface contribution develops more significantly for larger distances \cite{Wells1}. However, the observed four-point resistance does not approach the $1/s$ bulk behavior for $s \ge 50\,\mathrm{\mu m}$, because the space charge layer blocks the charge transport into the bulk due to the low conductivity of the depletion zone.
So, the four-point resistance in the equidistant range particularly reflects the properties of the space charge layer and the bulk, while the non-equidistant region is more suitable for the determination of the surface conductivity.
In total, the three-layer model including the intermediate layer describes the experimentally observed behavior very well.
Results obtained for other doping levels are shown in the Supplemental Material \cite{supmat} and confirm the results presented above.

An additional experiment is used to explore, if the bulk conductivity can be measured directly with the four-probe setup after removing the surface conductance channel. 
A hydrogen termination of the Si(111) surface by a treatment in HF, resulting in the formation of the Si(111)-(1$\times$1)-H, is known to remove the surface states present on the 7$\times$7 surface \cite{Chabal}.
The gray data points in Fig.~\ref{fig2} show the distance dependence of the four-point resistance in the non-equidistant region measured on a hydrogen-terminated Si(111) sample. 
The dashed green line corresponds to a fit using a pure 3D bulk behavior with a resistivity of $\rho_{3D} = (580 \pm 70)\,\mathrm{\Omega cm}$, which is close to the macroscopically measured nominal bulk resistivity of $(700 \pm 50)\,\mathrm{\Omega cm}$ and therefore confirms that without surface states a pure 3D bulk conductance is obtained. 

\begin{figure}[tbf]
\centering
\subfigure{\includegraphics[width=0.465\textwidth]{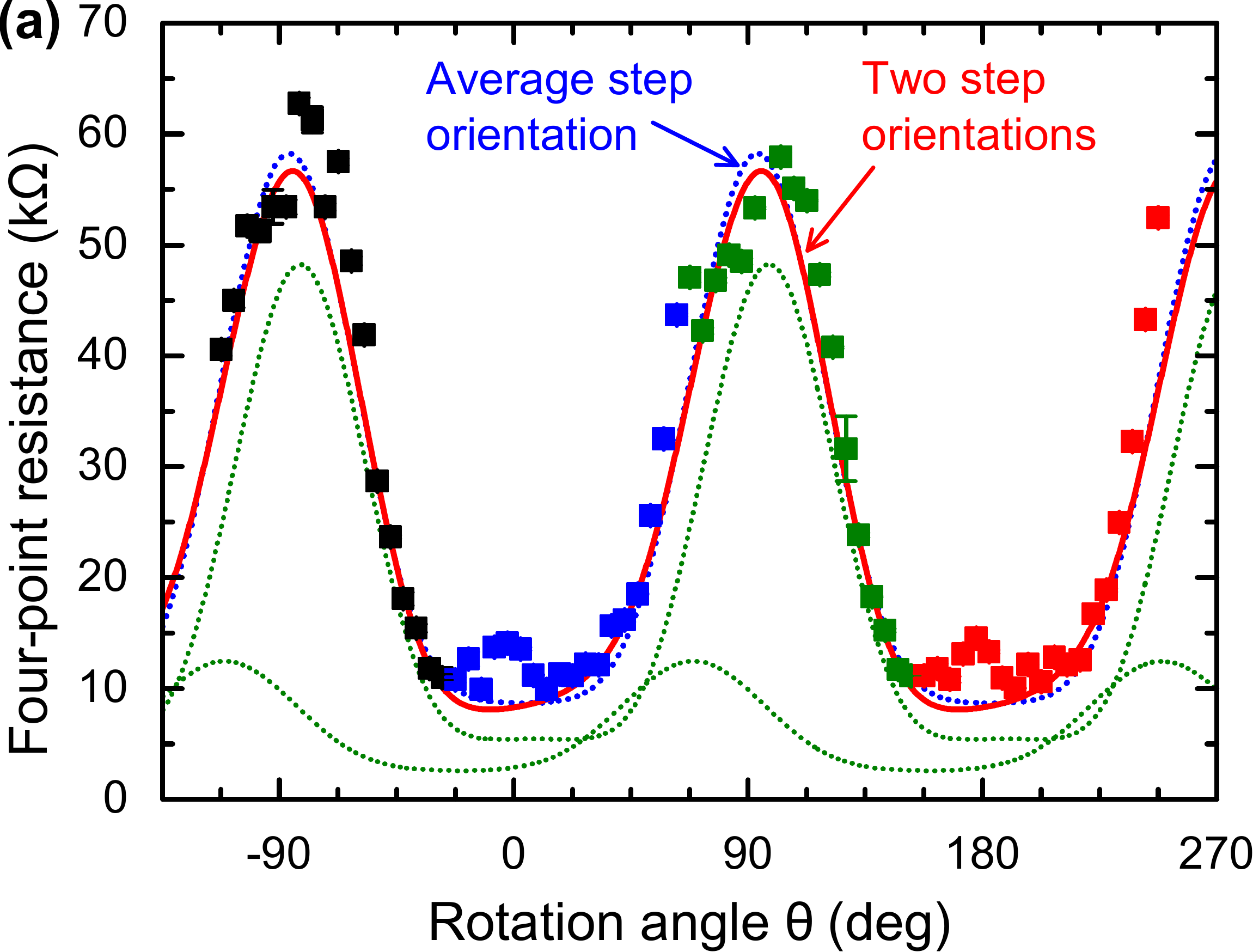}}\\
\subfigure{\includegraphics[width=0.459\textwidth]{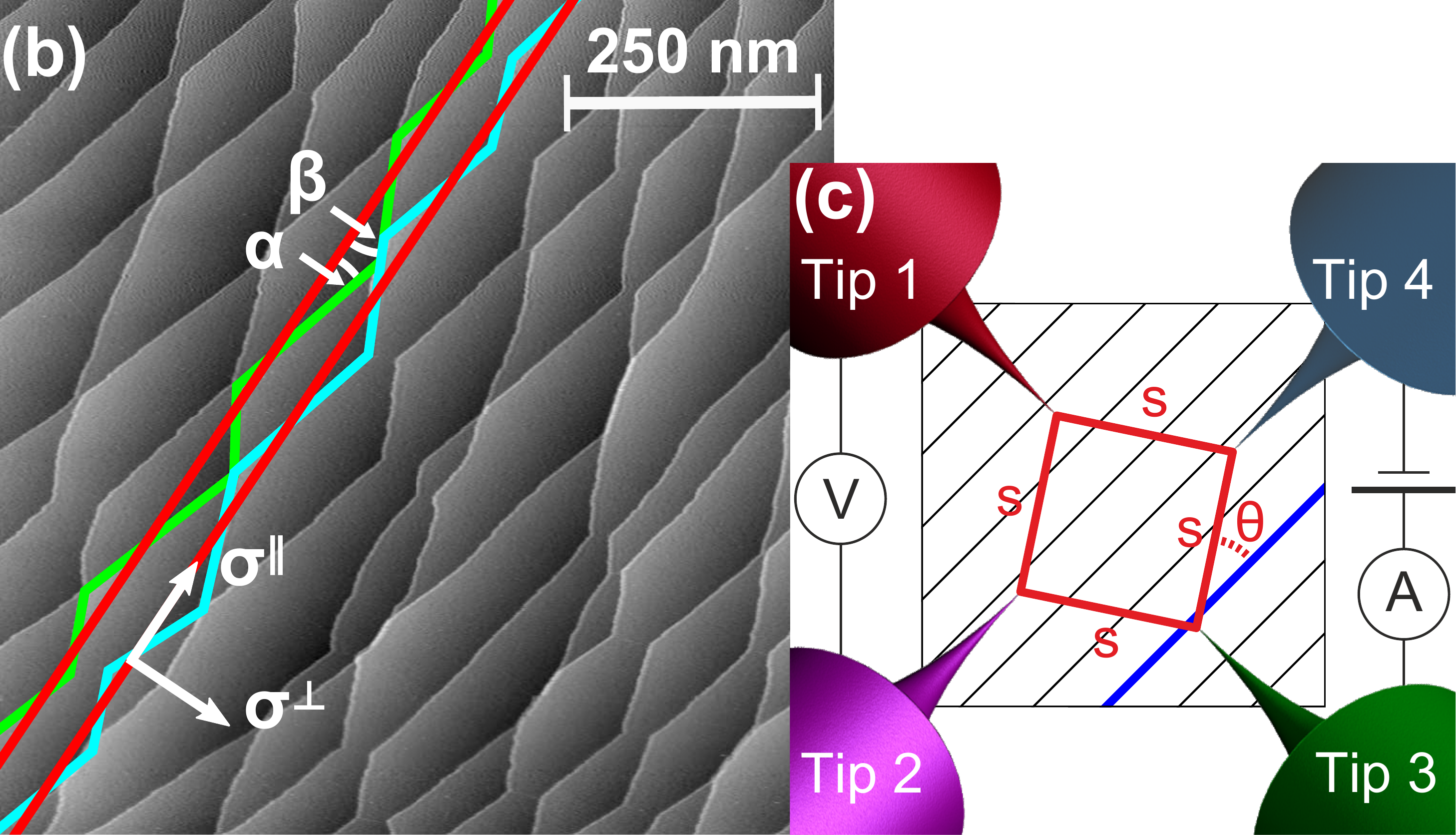}}%
\caption{(Color online) (a) Four-point resistance measured on a Si(111)-(7$\times$7) surface in the square configuration (shown in (c)) with $s = 50\,\mathrm{\mu m}$ as a function of rotation angle $\theta$ between the average step orientation and the line connecting the current injecting tips (colored data points). The fits to the data using either only one average step direction or a superposition of two step orientations (respective parts are shown as green dotted lines) are indicated as dotted blue line and solid red line, respectively. (b) STM image of the Si(111)-(7$\times$7) surface showing the representative step arrangement on the sample. Two adjacent step edges are highlighted (solid green and blue line) consisting of two main step directions indicated by the angles $\alpha$ and $\beta$ relative to the average step orientation (solid red lines).}
\label{fig3}
\end{figure}

While the distance-dependent four-point measurements could disentangle the surface conductivity from non-surface contributions to charge transport, the influence of atomic steps located on the (7$\times$7)-reconstructed Si surface has not been considered up to now. 
The conductivity arising from a single step for a current passing through it can be treated as scalar quantity. However, if a larger surface area is taken into account, 
the step array leads on average to an 
anisotropic conductivity described by the tensor components $\sigma^{\parallel}$ along the step edges and $\sigma^{\perp}$ perpendicular to the step edges \cite{Hasegawa2}. 
So, the anisotropic conductance is a macroscopic (mean field) result of the different number of step edges per unit length along different current paths. 
It turns out that the linear four-point measurement configuration (Fig.~\ref{fig1}) is not sensitive to a two-dimensional conductance anisotropy
\cite{Hasegawa3}. 
However, in a square arrangement of the four probes, as shown in Fig. \ref{fig3}(c), an angle-dependent four-point resistance is obtained from the solution of the Poisson equation for an anisotropic 2D sheet \cite{Tatarnikov,Hasegawa3} 
\begin{align}\label{eq:2}
R(\theta) = & \; C \cdot \ln\!\left(\!\!\frac{\left(\!\frac{\sigma^{\parallel}}{\sigma^{\perp}}+ 1 \!\right)^{\!2} - 4 \cos^2\!\theta \sin^2\!\theta \left(\!\frac{\sigma^{\parallel}}{\sigma^{\perp}} - 1\!\right)^{\!2}}{\left(\sin^2\!\theta + \frac{\sigma^{\parallel}}{\sigma^{\perp}}\cos^2\!\theta \right)^{\!2}}\!\right)
\end{align} 
with $C = 1/(4 \pi \sqrt{\sigma^{\parallel} \sigma^{\perp}})$.   

Results for the measured anisotropic four-point resistance on an n-doped Si(111)-(7$\times$7) sample ($700\,\mathrm{\Omega cm}$) for a probe spacing of $s = 50\,\mathrm{\mu m}$ are shown in Fig.~\ref{fig3}(a) as a function of the rotation angle $\theta$ relative to the step direction. 
The four sets of differently colored data points in angle increments of $5^\circ$ arise from the fact that for one fixed orientation of the probes four different rotation angles can be realized by successively assigning different probes as current and voltage probes. 

A fit of Eq.~\ref{eq:2} to the experimental data is shown as a dotted blue line in Fig.~\ref{fig3}(a) describing the angle dependence quite well \footnote{Since Eq.~\ref{eq:2} considers exclusively a 2D conductivity, a correction factor for the 6\% non-surface contribution to charge transport determined by the linear probe measurements has been taken into account.}. 
Nevertheless, the mean field approach applied so far assumes only straight step edges. 
However, the typical step structure present on our $0.25^\circ$ misoriented Si(111)-(7$\times$7) sample surface shown in Fig.~\ref{fig3}(b) consists of steps aligned mainly along two directions with average angles of $\alpha \approx 8^\circ$ and $\beta \approx 21^\circ$ with respect to the average step orientation (indicated as red lines), which now defines $\sigma^{\parallel}$ and $\sigma^{\perp}$.
This average step orientation arises from the macroscopic azimuthal direction of the sample miscut and is not aligned with the low-index orientations of the step edges. 
To model this more complicated non-parallel step pattern, we consider as first-order approximation a superposition of two step orientations with angles $\alpha$ and $\beta$ relative to the average step orientation weighted with their respective portion extracted from Fig.~\ref{fig3}(b). This leads to a slightly skewed curve shown as solid red line in Fig.~\ref{fig3}(a), which consists of an amount of 70\% and 30\% of the two single contributions, respectively (dotted green lines). 
The model including the two step orientations describes the data as well as Eq. \ref{eq:2}, but contains a better approximation of the sample step structure, 
and results in 
$\sigma_{\parallel} = (9 \pm 2)\cdot 10^{-6}\,\mathrm{\Omega}^{-1}/\square$ and $\sigma_{\perp} = (1.7 \pm 0.4)\cdot10^{-6}\,\mathrm{\Omega}^{-1}/\square$ with an anisotropy ratio of $\sigma_{\parallel}/\sigma_{\perp} \approx 5$. 
The geometric mean $\sqrt{\sigma^{\parallel} \sigma^{\perp}} = (3.9 \pm 0.6) \cdot 10^{-6}\,\mathrm{\Omega}^{-1}/\square$ has nearly the same value within the error tolerances as the surface conductivity $\sigma_{2D}$ obtained in the linear configuration. Thus, the two independent methods, the distance-dependent linear configuration and the angle-dependent square configuration, yield the same results for the surface conductivity. 

In a last step, we approximate the measured mean field anisotropic conductivity by the scalar resistivities of a step-free terrace $\rho_{\mathrm{surf}}$ and a single step $\rho_{\mathrm{step}}$. Considering first the direction parallel to the steps, no step edges have to be crossed by the current, which results in the relation 
\begin{align}
	1/\sigma^{\parallel} & = \rho^{\parallel} = \rho_\mathrm{surf} \label{eq:3-1} \, \mathrm{.} 
\end{align}
Second, the resistivity perpendicular to the step edges is composed of additive contributions from the steps and the step-free terraces and can be expressed as series resistance, resulting in \cite{Hasegawa2} 
\begin{align}
	1/\sigma^{\perp} & = \rho^{\perp} = \rho_\mathrm{surf} + \rho_\mathrm{step}/d_\mathrm{step}^{\perp} \label{eq:3-2}
\end{align}
with $d_\mathrm{step}^{\perp}$ denoting the average distance between the steps. From the two relations in Eq.~\ref{eq:3-1} and Eq. \ref{eq:3-2} finally the conductivity of the step-free Si(111)-(7$\times$7) surface can be disentangled from the influence of the step conductivity as 
$\sigma_\mathrm{surf} = (9 \pm 2) \cdot 10^{-6}\,\mathrm{\Omega}^{-1}/\square$, and $\sigma_\mathrm{step} = (29 \pm 9)\,\mathrm{\Omega}^{-1} \mathrm{m}^{-1}$. The value of the surface conductivity $\sigma_\mathrm{surf}$ is a factor of $2$ to $6$ larger than the values obtained in recent experiments \cite{Hasegawa1,Wolkow}. 
Such smaller values may be explained, as these experiments are based on a more indirect comparison of the conductivity before and after quenching the surface states by adsorption of atoms/molecules. 
For the quenched system several conditions have to be fulfilled: (a) the surface sates of the surface under study are completely quenched, (b) the space charge layer conductivity is not influenced by the adsorbed layer, and (c) the adsorbed layer induces no (additional) surface conductance. If one of these conditions is not fulfilled, these experiments based on the difference method result in different values for the surface conductivity. 

From a comparison of the surface resistivity and the step resistivity, the following relation is obtained. The resistance of one step (per unit length) corresponds to the resistance of a segment of the step-free Si(111)-(7$\times$7) surface (per unit length) of a width of $300 \,\mathrm{nm}$. For our sample with a step density of $14\,\mathrm{steps}/\mathrm{\mu m}$, the contribution of the step resistance to the total resistance has a substantial amount 
of 80 \% for a current flowing in the perpendicular direction. In general, the presence of steps will reduce the surface conductivity of the Si(111)-(7$\times$7) considerably, however, in a well predictable manner. 

In conclusion, we combined the distance-dependent linear configuration for four-point resistance measurements on Si(111) surfaces with a three-layer model for charge transport in order to disentangle the surface conductivity from non-surface contributions (bulk and space charge layer conductivity).  
The influence of atomic surface steps is obtained by measurements in the angle-dependent square configuration resulting in a step-free surface resistivity of $\rho_{\mathrm{surf}} = (116 \pm 26)\,\mathrm{k \Omega}/\square$ and a step resistivity of $\rho_{\mathrm{step}} = (3.4 \pm 1) \cdot 10^{-2}\,\mathrm{\Omega\,m}$ for the Si(111)-(7$\times$7) surface.  
These two generic methods can easily be used to determine surface conductivities of other mixed 2D/3D systems, like different semiconductors or topological insulators.

\balancecolsandclearpage


\setcounter{equation}{0}
\setcounter{figure}{0}
\setcounter{table}{0}
\setcounter{page}{1}
\makeatletter
\renewcommand{\theequation}{S\arabic{equation}}
\renewcommand{\thefigure}{S\arabic{figure}}
\renewcommand{\bibnumfmt}[1]{[S#1]}
\renewcommand{\citenumfont}[1]{S#1}
\makeatother
\widetext 

\section{\noindent \textbf{Supplemental Material for "Surface and Step Conductivities on Si(111) Surfaces", by S. Just, M. Blab, S. Korte, V. Cherepanov, H. Soltner, and B.~Voigtl\"{a}nder}*}

In this supplemental material, details about sample preparation and measurement procedure are described, and some additional experimental results obtained for differently doped Si samples are presented. Furthermore, the three-layer model used to describe the four-point resistance measured in the distance-dependent linear probe configuration is discussed. 

\section{Details of the sample preparation and measurement procedure}

The sample preparations and measurements are carried out under ultra-high vacuum (UHV) conditions with a base pressure of $\sim 1 \cdot 10^{-10}\,\mathrm{mbar}$. Subsequent to the cleaning process by direct current heating to $1230\,^{\circ}\mathrm{C}$ the Si(111) samples are cooled down slowly, especially in the vicinity of the transition temperature of $\sim 800\,^{\circ}\mathrm{C}$, for establishing the 7$\times$7 surface reconstruction. 
%
%
%
%
%
Bi passivation is obtained afterwards as described in \cite{s:Romanyuk} to achieve a $(\sqrt{3} \times \sqrt{3})$R30$^\circ$ surface terminated with 1\,ML bismuth.  
For the H-termination, resulting in the formation of a Si(111)-(1$\times$1)-H surface, a treatment in a 1\% solution of HF acid is used, and after an additional cleaning step in deionized water the sample is transferred to the ultra-high vacuum within a time of 10 minutes. 

The distance-dependent four-point measurements are performed in a four-tip scanning tunneling microscope \cite{s:4tipSTM}. In this system the individual positioning of the tips is realized under the control of an optical microscope. This has the advantage compared to an electron microscope that frequently observed influences of the electron beam on the surface properties are avoided \cite{s:Uetake,s:Lepsa,s:Berthe}. 
In the linear arrangement of the four probes the voltage between the inner two tips is measured as a function of the current injected by the outer tips. For the square configuration, the voltage probes and current probes are located at the corners of opposite sides of the square, and by successively rotating the assignment of these probes different rotation angles in multiples of $90^{\circ}$ can be realized for one fixed position of the tips.  
%
%
%
The four-point resistance $R^{4p}$ is obtained from the slope of the measured I/V curves close to zero volt, whereby each resistance value is averaged over four I/V curves. 

As for the four-point measurements preferentially a non-equidistant probe spacing is used, in which three tips remain at a distance $s$ and only the distance $x$ between one outer current injecting tip and the adjacent voltage measuring tip is varied, Eq. 1 in the main article has to be modified in this case to \cite{s:Wells2,s:Wells3,s:Wojtaszek}\\[-1ex] 
\begin{equation}\label{s:eq:1-2}
	R^{4p}_{2D}= \frac{1}{\chi_{2D}} \cdot \frac{\ln 2}{\pi \sigma_{2D}}, \quad \mathrm{and} \quad R^{4p}_{3D}=\frac{1}{2 \pi \sigma_{3D}} \cdot {s^{eq.}_{3D}}^{-1}
\end{equation}
with the \textit{2D sensitivity} $\chi_{2D}$
\begin{equation}\label{s:eq:1-2-1}
	\frac{1}{\chi_{2D}}(s,x) = \frac{1}{2 \ln 2} \cdot \left(\ln \frac{2s}{x} + \ln \frac{s+x}{s} \right) 
\end{equation}
and the \textit{effective spacing} $s^{eq.}_{3D}$
\begin{equation}\label{s:eq:1-2-2}
	\frac{1}{s^{eq.}_{3D}}(s,x) = \frac{1}{x} + \frac{1}{2s} - \frac{1}{s+x} \; \mathrm{.}\\[2ex]
\end{equation}

For contacting the sample surface, in a first step the four tips are approached until a tunnel contact with a current in the low nA range is established. Then, after retracting the tips by several nm, the feedback of the STM is switched off, and the tips are manually approached further separately until an increase in the current up to 1 $\mu$A is observed. At this point, the tips are in contact with the sample, but they only touch the surface and have a penetration depth of only a few~$\mathrm{\AA}$ as confirmed by separate experiments. For transport measurements, the sample is set to floating potential and a current is injected by an applied voltage ramp between the outer tips. Since the voltage at the inner tips is measured as function of the actual current, there is no influence of a potentially fluctuating contact resistance and resulting variation in current through the sample due to the switched-off feedback.

\section{Additional experimental results}

\begin{figure}[t!]
\centering
\subfigure{\includegraphics[width=0.49\textwidth]{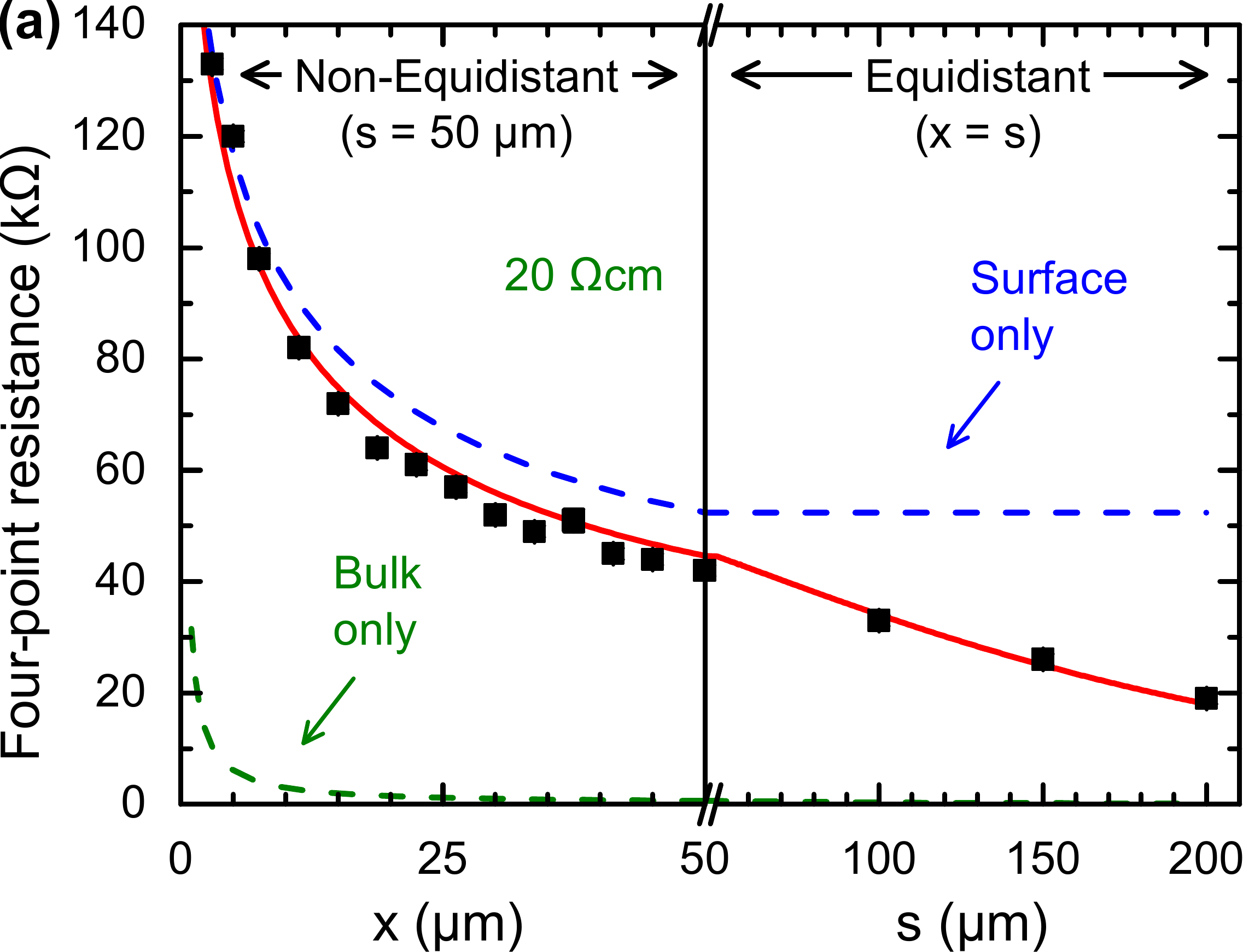}}\\
\subfigure{\includegraphics[width=0.49\textwidth]{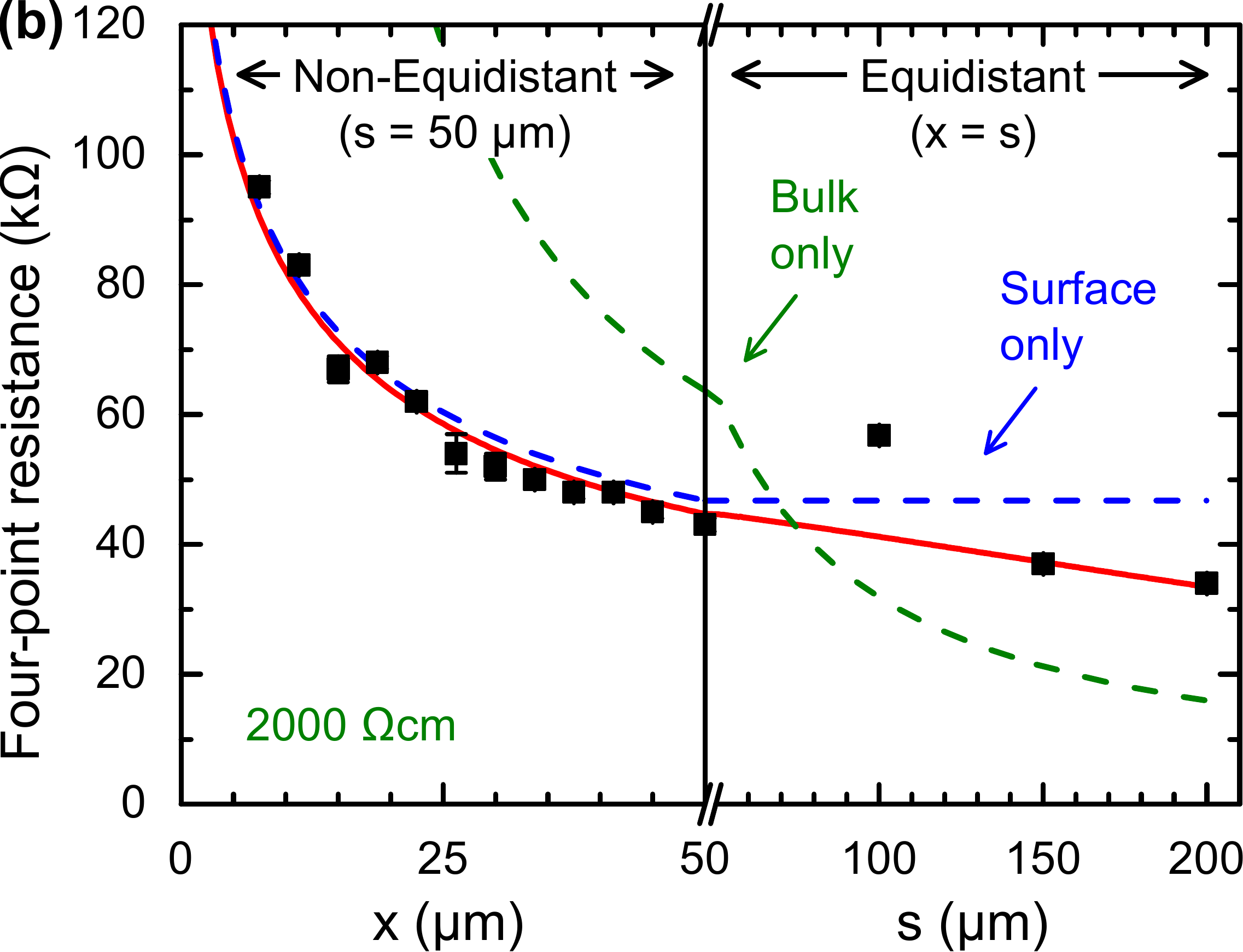}}\\
\subfigure{\includegraphics[width=0.49\textwidth]{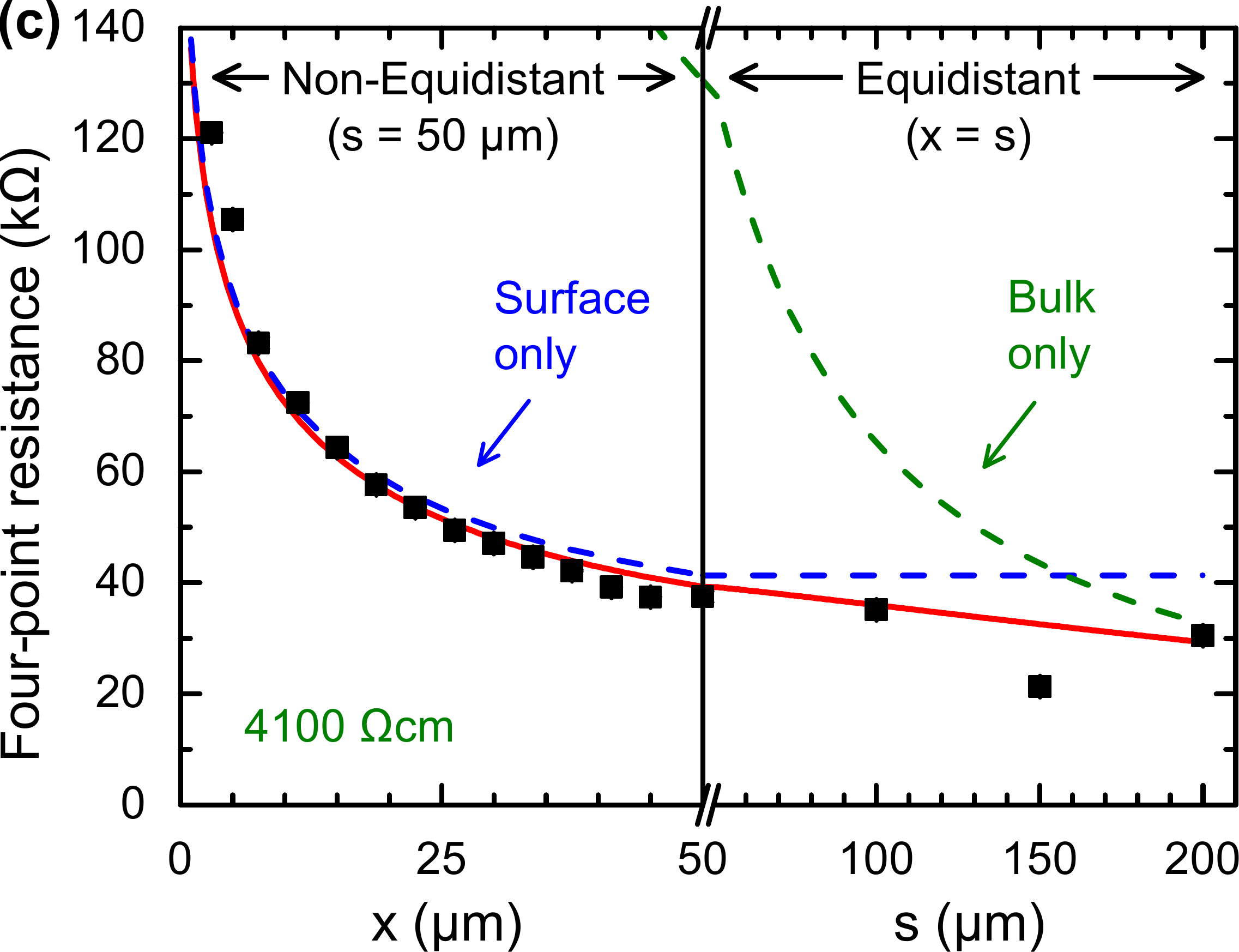}}%
\caption{Measured four-point resistance of n-doped Si(111)-(7$\times$7) samples with bulk resistivities of (a) $20\,\Omega\mathrm{cm}$, (b) $2000\,\Omega\mathrm{cm}$ and (c) $4100\,\Omega\mathrm{cm}$ as function of the probe distances $s$ and $x$ for the equidistant configuration as well as the non-equidistant configuration with $ s = 50\,\mathrm{\mu m}$. The decreasing four-point resistance in the equidistant region indicates a non-surface conductance contribution due to bulk and space charge layer. Applying a three-layer model for transport in the sample yields the solid red line located between the two limiting cases of a pure surface conductance  (dashed blue line) and a pure bulk conductance (dashed green line).}
\label{s:fig1}
\vspace{-0.95cm}
\end{figure}

Additional four-point resistance measurements in the linear probe configuration are carried out on differently n-doped Si(111)-(7$\times$7) substrates with bulk resistivities ranging over two orders of magnitude. The results for the non-equidistant probe spacing with $x \le 50\,\mathrm{\mu m}$ and $s = 50\,\mathrm{\mu m}$, as well as the equidistant spacing with distances $x = s \ge 50\,\mathrm{\mu m}$ are shown in Fig. \ref{s:fig1} for three Si samples with resistivities of (a) $20\,\Omega\mathrm{cm}$, (b) $2000\,\Omega\mathrm{cm}$ and (c) $4100\,\Omega\mathrm{cm}$.
From the best fits according to the used three-layer model (solid red line), the surface conductivities for the three different samples can be determined to $\sigma_{(a)}\,=\,(4.2 \pm 0.6)\cdot 10^{-6}\,\mathrm{\Omega}^{-1}/\square$, $\sigma_{(b)}\,=\,(4.7 \pm 0.6)\cdot 10^{-6}\,\mathrm{\Omega}^{-1}/\square$ and $\sigma_{(c)}\,=\,(5.3 \pm 0.8)\cdot 10^{-6}\,\mathrm{\Omega}^{-1}/\square$. As the measured surface conductivity in the linear configuration is a combination of contributions due to step edges and the step-free surface, these values can differ slightly from each other for the three differently doped Si samples because of slightly different miscut angles of the substrates and resulting different step densities at the surfaces. Nevertheless, all obtained surface resistivities (including the $700\,\Omega\mathrm{cm}$ sample discussed in the main text with $\sigma\,=\,(5.1 \pm 0.7)\cdot 10^{-6}\,\mathrm{\Omega}^{-1}/\square$) are very close to each other and still compatible within the error tolerances indicating that the step contributions are similiar for all samples.
Comparing the measured data to the two limiting cases of a pure 2D conductivity with the above values and a pure 3D conductivity arising only from the bulk shown as dashed blue and green lines in Fig. \ref{s:fig1}, respectively, one can see that the observed resistance behavior is very close to the 2D case for all of the three differently doped Si(111) samples, although the bulk resistance varies over several orders of magnitude. 
This indicates a separation of the surface layer from the bulk arising from the space charge region with low conductivity due to a depletion zone preventing an enhanced charge transport through the bulk, which especially becomes important for high bulk doping concentrations. 
For larger probe distances, the measured four-point resistance increasingly deviates from the 2D case and shows a pronounced non-surface contribution resulting from additional charge transport through space charge layer and bulk. This non-surface contribution decreases for lower bulk doping levels, as it is expected in comparison with the large increase in bulk resistance in the limiting case of a pure 3D conductivity. 
Within the three-layer model, the space charge region is approximated by one intermediate layer with thicknesses of (a) $0.9\,\mathrm{\mu m}$, (b) $4.8\,\mathrm{\mu m}$ and (c) $5.3\,\mathrm{\mu m}$, and with constant conductivities of (a) $1.5 \cdot 10^{-4}\,\mathrm{\Omega}^{-1} \mathrm{m}^{-1}$, (b) $1.8 \cdot 10^{-4}\,\mathrm{\Omega}^{-1} \mathrm{m}^{-1}$ and (c) $3.2 \cdot 10^{-4}\,\mathrm{\Omega}^{-1} \mathrm{m}^{-1}$, respectively, for the three differently doped samples in Fig. \ref{s:fig1}(a)--(c).

\section{Three-layer model for conductance}

\begin{figure}[t]
	\centering
		\includegraphics[width=0.3\textwidth]{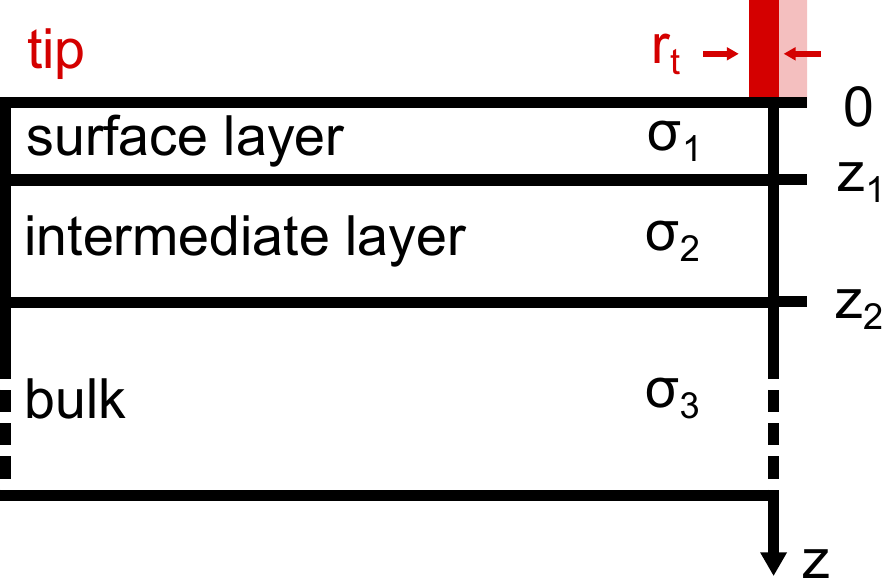}
		\caption{The three-layer model consists of a thin surface layer, an intermediate layer and the semi-infinite bulk described by their respective conductivities $\sigma_1$, $\sigma_2$ and $\sigma_3$, and their positions $z_1$ and $z_2$. The current $I$ is injected by a cylindrical tip of radius $r_t$ at the surface layer. }
	\label{s:fig2}	
\end{figure}

The three-layer model assumes a layered sample structure shown in Fig. \ref{s:fig2} consisting of a thin surface layer, an intermediate layer and a semi-infinite bulk characterized by their respective conductivities $\sigma_1$, $\sigma_2$ and $\sigma_3$, and positions of the interfaces $z_1$ and $z_2$.  At the surface a current $I$ is injected by a cylindrical tip with radius $r_t$. Due to calculation requirements, the surface layer cannot be two-dimensional, so that a finite thickness of one atomic layer ($3\, \mathrm{\AA}$) is assumed. As $\nabla \cdot \mathbf{j} = 0$ for the current density $\mathbf{j} = \sigma \mathbf{E} = -\sigma\, \nabla \Phi$ inside the sample (excluding the injection point), the electrical potential~$\Phi$ in this region can be determined by solving the Laplace equation  
\begin{align}
	\Delta \Phi & = 0
\end{align}
in cylindrical coordinates. Taking account of the angle-independent polar symmetry for one tip, a solution for the potential in the individual layers is \cite{s:Jackson} 
\begin{align}
	\Phi_1(\rho,z) & = \int_0^\infty \left[a(k) \, e^{kz} + b(k) \,e^{-kz}\right] J_0(k\rho) \, \mathrm{d}k \\[0.5ex] 
	\Phi_2(\rho,z) & = \int_0^\infty \left[c(k) \, e^{kz} + d(k) \,e^{-kz}\right] J_0(k\rho) \, \mathrm{d}k \\[0.5ex]
	\Phi_3(\rho,z) & = \int_0^\infty f(k) \,e^{-kz} \, J_0(k\rho) \, \mathrm{d}k  
\end{align}
with $J_0$ denoting the Bessel function of the first kind. The corresponding boundary conditions are 
\begin{align}
	\sigma_1 \frac{\partial}{\partial z}\Phi_1(\rho,0) & = - j_1\,  H(r_t - \rho) \label{s:bound1}\\[0.5ex]
	\Phi_1(\rho,z_1) &= \Phi_2(\rho,z_1) \label{s:bound2}\\[0.5ex]
	\sigma_1 \frac{\partial}{\partial z}\Phi_1(\rho,z_1) & = \sigma_2 \frac{\partial}{\partial z}\Phi_2(\rho,z_1) \label{s:bound3}\\[0.5ex]
	\Phi_2(\rho,z_2) &= \Phi_3(\rho,z_2) \label{s:bound4}\\[0.5ex]
	\sigma_2 \frac{\partial}{\partial z}\Phi_2(\rho,z_2) & = \sigma_3 \frac{\partial}{\partial z}\Phi_3 (\rho,z_2) \label{s:bound5}
\end{align}   
resulting from the current injection (Eq.~\ref{s:bound1}), as well as from the continuous transitions of the potential (Eq.~\ref{s:bound2} and Eq.~\ref{s:bound4}) and the current density (Eq.~\ref{s:bound3} and Eq.~\ref{s:bound5}) between the layers. In Eq. \ref{s:bound1}, the expression $H(r_t - \rho)$ denotes the Heaviside step function. The injected current density can be described by $j_1 = \frac{I}{\pi\,r_t^2}$ assuming a cylindrical tip with a tip radius of $r_t \approx 25\,  \mathrm{nm}$, which seems reasonable for an STM tip. Nevertheless, it turns out that also other values for the tip radius in the range of $5\,\mathrm{nm}$ to $100\,\mathrm{nm}$ do not influence the results of the calculations in a considerable manner. Based on Eqs.~\ref{s:bound1} $-$ \ref{s:bound5}, a matrix equation determining the coefficients $a(k),\dotsc, f(k)$ is derived
\begin{align}
	\begin{pmatrix}
		1 & -1 & 0 & 0 & 0 \\[1ex]
		e^{k z_1} & -e^{-k z_1} & - \frac{\sigma_2}{\sigma_1}\, e^{k z_1} & \frac{\sigma_2}{\sigma_1}\, e^{-k z_1} & 0 \\[1ex]
		e^{k z_1} & e^{-k z_1 } & - e^{k z_1} & -e^{-k z_1 } & 0 \\[1ex]
		0 & 0 & e^{k z_2 } & -e^{-k z_2} & \frac{\sigma_3}{\sigma_2}\, e^{-k z_2}\\[1ex]
		0 & 0 & e^{k z_2} & e^{-k z_2} & -e^{-k z_2} 
	\end{pmatrix}
	\cdot %
	\begin{pmatrix}
		a(k)\\[1ex]
		b(k)\\[1ex]
		c(k)\\[1ex]
		d(k)\\[1ex]
		f(k)
	\end{pmatrix}
	& = %
	\begin{pmatrix}
		- \frac{j_1}{\sigma_1} \, \int_0^{r_{t}} \rho \, J_0(k\rho)\,\mathrm{d}\rho  \\[1ex]
		0\\[1ex]
		0\\[1ex]
		0\\[1ex]
		0
	\end{pmatrix}\,\mathrm{,} \label{s:matrix}
\end{align}
which can be solved by means of matrix inversion. As the potential at the surface ($z = 0$) can be expressed by 
\begin{align}
	\Phi_{\mathrm{surf}}(\rho) = \Phi_1(\rho,0) & = \int_0^\infty \left[a(k) + b(k)\right] J_0(k\rho) \, \mathrm{d}k \,\mathrm{,}
\end{align} 
only the coefficients $a(k)$ and $b(k)$ are relevant for the calculation. From Eq. \ref{s:matrix} an expression for the sum of these coefficients is obtained 
\begin{align}
	a(k) + b(k) & = - \frac{j_1}{\sigma_1} \, \int_0^{r_{t}} \rho \, J_0(k\rho)\,\mathrm{d}\rho  \cdot \frac{\frac{\frac{\sigma_3}{\sigma_2} \tanh\left[k(z_2 - z_1)\right] + 1}{\tanh\left[k(z_2 - z_1)\right] + \frac{\sigma_3}{\sigma_2}} + \frac{\sigma_2}{\sigma_1} \tanh\left[kz_1\right]}{\frac{\frac{\sigma_3}{\sigma_2} \tanh\left[k(z_2 - z_1)\right] + 1}{\tanh\left[k(z_2 - z_1)\right] + \frac{\sigma_3}{\sigma_2}} \cdot \tanh\left[kz_1\right] + \frac{\sigma_2}{\sigma_1} }\;\mathrm{.}
\end{align} 
Introducing cartesian coordinates with $\mathbf{x} = \begin{pmatrix} x & y \end{pmatrix}^T $ and $\rho = |\mathbf{x} - \mathbf{x_{0}}| = \sqrt{(x-x_{0})^2+(y-y_{0})^2}$ for a tip positioned at $\mathbf{x_0}$, the combined potential on the surface $\Phi_{\mathrm{surf},12}$ for a current source at position $\mathbf{x_{0_1}}$ and a current sink at position $\mathbf{x_{0_2}}$ results in
\begin{align}
	\Phi_{\mathrm{surf},12}(\mathbf{x}) & = \Phi_{\mathrm{surf},1}(|\mathbf{x} - \mathbf{x_{0_1}}|) -  \Phi_{\mathrm{surf},2}(|\mathbf{x} - \mathbf{x_{0_2}}|) \,\mathrm{.} 
\end{align}
Finally, the four-point resistance $R^{4p}$ measured on the surface is determined by the quotient of the potential difference between the positions $\mathbf{x_{0_3}}$ and $\mathbf{x_{0_4}}$ of the measuring tips, and the current $I$ resulting in 
\begin{align}
	R^{4p} & = \frac{\Phi_{\mathrm{surf},12}(\mathbf{x_{0_3}}) - \Phi_{\mathrm{surf},12}(\mathbf{x_{0_4}})}{I} \\[2.25ex]
	       & \!\begin{multlined}[b][0.575\displaywidth]
	       = \frac{1}{I} \int_0^\infty \left[ a(k) + b(k) \right] \cdot \left[ J_0(k\, |\mathbf{x_{0_3}} - \mathbf{x_{0_1}}|) -  J_0(k\, |\mathbf{x_{0_3}} - \mathbf{x_{0_2}}|) \right. \\ 
	       \left. - J_0(k\, |\mathbf{x_{0_4}} - \mathbf{x_{0_1}}|)  + J_0(k\, |\mathbf{x_{0_4}} - \mathbf{x_{0_2}}|)  \right] \, \mathrm{d}k\, \mathrm{.}  \label{s:r4p}
\end{multlined} 
\end{align}
For the linear probe configuration with equidistant spacing $s$ between three tips and a non-equidistant spacing $\tilde x$ between one outer current tip and the adjacent voltage measuring tip, Eq. \ref{s:r4p} simplifies to 
\begin{align}
	R^{4p}(s,\tilde x)  & = \frac{1}{I} \int_0^\infty \left[ a(k) + b(k) \right] \cdot \left[ J_0(k s) - J_0(k (s  + \tilde x )) - J_0(2 k s) + J_0(k \tilde x)  \right] \, \mathrm{d}k \, \mathrm{.} \label{s:r4p-linear}
\end{align}

So, Eq. \ref{s:r4p-linear} describes the equidistant measurement range according to Fig. \ref{s:fig1}(a)--(c), if $\tilde x = s  \ge 50 \,\mu\mathrm{m}$, and the non-equidistant region, if $\tilde x \le s = 50 \,\mu\mathrm{m}$ is assumed. The integral over the Bessel functions can be evaluated numerically and the result can be fitted to the measurement data with the free parameters $\sigma_1$, $\sigma_2$ and $z_2$ defining the properties of the surface layer and the intermediate layer, which is used as approximation for the space charge region in the sample. The value for $\sigma_3$ is known from macroscopic measurements of the bulk resistivity and is in agreement with the nominal doping concentration.  

\begin{figure}[t]
\centering
\includegraphics[width=0.51\textwidth]{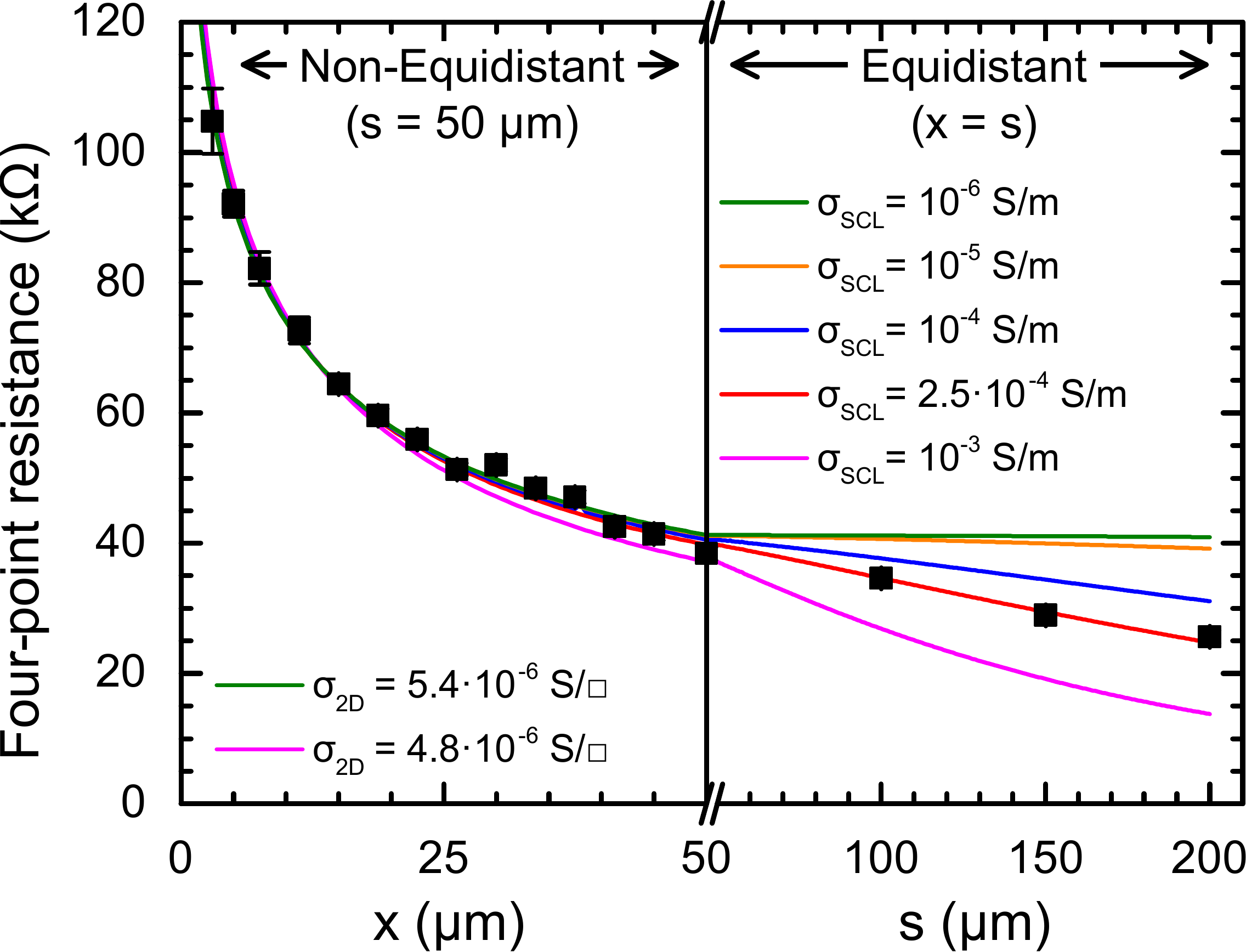}
\caption{Description of the measured four-point resistance (n-doped Si(111)-(7$\times$7) sample, bulk resistivity $700\,\Omega\mathrm{cm}$) by the three-layer model for different input parameters (colored curves). For the nominal bulk conductivity and a space charge layer conductivity varying over three orders of magnitude from $1 \cdot 10^{-3}\,\mathrm{\Omega}^{-1} \mathrm{m}^{-1}$ (magenta curve) to $1 \cdot 10^{-6}\,\mathrm{\Omega}^{-1} \mathrm{m}^{-1}$ (green curve) the measured data are fitted with the surface conductivity as free paramter. All curves can describe the data in the region of small probe spacings ($x \le 50\,\mathrm{\mu m}$) very well, while there are considerable deviations from the data in the region of larger probe distances ($s \ge 50\,\mathrm{\mu m}$). Nevertheless, the obtained surface conductivity shows only a very minor spread and varies less than 10\% from $\sigma_{2D}\,=\,4.8 \cdot 10^{-6}\,\mathrm{\Omega}^{-1}/\square$ (magenta curve) to $\sigma_{2D}\,=\,5.4 \cdot 10^{-6}\,\mathrm{\Omega}^{-1}/\square$ (green curve). This indicates that the measured four-point resistance in the small probe spacing region does not depend on the details of the space charge layer, and so the surface conductivity can be determined very precisely, even if the exact properties of the space charge layer are not known.}
%
\label{s:fig5}
\end{figure}

The surface conductance channel has the largest influence on the measured four-point resistance for small distances, i.e. $\tilde x \le 50\,\mathrm{\mu m}$ (region of non-equidistant probe spacing), while the properties of the intermediate layer (conductivity and thickness) have the largest influence on the four-point resistance in the region of larger distances (equidistant spacing region). This is illustrated in Fig.~\ref{s:fig5}, in which the measured four-point resistance (n-doped Si(111)-(7$\times$7) sample, bulk resistivity $700\,\Omega\mathrm{cm}$) is compared to theoretical curves resulting from the three-layer model for different input parameters (colored curves). For the constant nominal bulk resisitivity of the sample and a varying conductivity for the intermediate layer (space charge layer) over three orders of magnitude from $1 \cdot 10^{-3}\,\mathrm{\Omega}^{-1} \mathrm{m}^{-1}$ (magenta curve) to $1 \cdot 10^{-6}\,\mathrm{\Omega}^{-1} \mathrm{m}^{-1}$ (green curve), the measured data are fitted to determine the surface conductivity. 
If the space charge layer conductivity is enhanced above a value of $1 \cdot 10^{-3}\,\mathrm{\Omega}^{-1} \mathrm{m}^{-1}$, the measured data cannot be described any more by the model and a fit is not possible. So, the space charge layer should be described by a conductivity value in the chosen region.  
Although there is a very large spread in the space charge layer conductivity, the results show a very minor spread of the surface conductivity with a deviation of less than 10\% from $\sigma_{2D}\,=\,4.8 \cdot 10^{-6}\,\mathrm{\Omega}^{-1}/\square$ (magenta curve) to $\sigma_{2D}\,=\,5.4 \cdot 10^{-6}\,\mathrm{\Omega}^{-1}/\square$ (green curve). 
All fit curves describe the data very well in the region of small probe spacing ($x \le 50\,\mathrm{\mu m}$), while there are considerable deviations from the data points for larger probe spacing ($s \ge 50\,\mathrm{\mu m}$). This confirms the influence of surface conductivity and space charge layer conductivity in different regions of the used probe distances. Thus the surface conductivity can be determined very precisely from the non-equidistant probe spacing region, even if the properties of the space charge layer are taken into account only very approximately. On the other hand, the parameters of the space charge layer can be further approximated from the measurement data in the equidistant region. 
%
%
In total, this justifies the crude approximation of the space charge region as only one layer with constant conductivity, as the quantity to be determined, the surface conductivity $\sigma_1$, does not depend significantly on the values of the conductivity $\sigma_2$ and width $z_2$ used to describe the intermediate layer.

The three-layer model described above can easily be extended to a multi-layer model consisting of $N$ separate layers. In this case, the boundary conditions in Eqs.~\ref{s:bound1} -- \ref{s:bound5} have to be modified to include the transition between the layer $n-1$ and layer $n$ for $n = 1,\dotsc,(N-1)$ and the size of the matrix in Eq. \ref{s:matrix} becomes $(2N-1)\,\times\,(2N-1)$ requiring a numerical solution method, but Eq. \ref{s:r4p} for the four-point resistance remains the same. So, if a depth-dependent conductivity profile is known, the charge transport in a sample can be described very precisely. 

In contrast to the analytical model, one other way to simulate the contributions of the different charge transport channels to the four-point resistance is to invoke finite element calculations. We have not done this, since the distances involved range from a few nanometers (radius of the current injection) to $200\,\mathrm{\mu m}$ (maximum probe distance) and such a large range of length scales is difficult to include in finite element calculations. Nevertheless, as the exact analytical solution of the potential problem in a layered sample has quite an elementary form, the three-layer model is easier to apply and provides a more accurate computation than a finite element simulation.

\end{document}